\documentclass[12pt]{article}

\usepackage{amsmath,amssymb}
\usepackage{amscd}

\newcommand{\Rr}{\mathbb{R}}
\newcommand{\Ss}{\mathbb{S}}
\newcommand{\CC}{\mathbb{C}}
\newcommand{\HH}{\mathcal{H}}
\newcommand{\Hh}{\mathfrak{H}}
\newcommand{\Dd}{\mathcal{D}}

\DeclareMathOperator{\tr}{tr}
\DeclareMathOperator{\ch}{ch}
\DeclareMathOperator{\sh}{sh}

\begin{document}

\begin{center}
\Large{Zero modes for the quantum Liouville model}
\vspace{0.5cm}

\large{L.~D.~Faddeev\footnote{The work was partially supported by RFBR grants 11-01-00570-a, 11-01-12037-ofi-m and the programm
<<Mathematical problems of nonlinear dynamics>> of Russian Academy of Science.}}\\
\vspace{0.5cm}
    \emph{St.~Petersburg Dept. of Steklov Mathematical Institute}\\
    and\\
    \emph{St.~Petersburg State University}\\
\end{center}

\begin{abstract}
    The problem of definition of zero modes for quantum Liouville model
    is discussed and corresponding Hilbert space representation
    is constructed.
\end{abstract}

\section{Introduction and statement of the problem}
    
    Quantum Liouville model is a basic example of conformal field theory.
    It represents a quantization of the dynamical system for real-valued
    field
$ \phi(x,t) $,
    defined on the cylinder
$ (x,t) \in \Ss^{1}\times \Rr $
    by hamiltonian
\begin{equation*}
    H = \frac{1}{2\gamma} \int_{0}^{2\pi} \bigl(\pi^{2} + \phi_{x}^{2}
	+ e^{2\phi} \bigr) dx
\end{equation*}
    and canonical Poisson bracket
\begin{equation*}
    \{\phi(x), \phi(y)\} = 0 , \quad
    \{\pi(x), \pi(y)\} = 0 , \quad
    \{\pi(x), \phi(y)\} = \gamma \delta(x-y)
\end{equation*}
    in terms of the initial data
$ \phi(x) = \phi(x,0) $,
$ \pi(x) = \phi_{t}(x,0) $.
    The corresponding equation of motion looks as follows
\begin{equation}
\label{Leqm}
    \phi_{tt} - \phi_{xx} + e^{2\phi} = 0 .
\end{equation}
    The quantization developed by several authors
(see survey \cite{Gervais} and the literature cited there)
    is based on the classical Liouville formula
\begin{equation}
\label{Lf}
    e^{2\phi} = -4 \frac{f'(x-t)g'(x+t)}{\bigl(f(x-t) - g(x+t)\bigr)^{2}} ,
\end{equation}
    and parametrization of functions
$ f(x) $ and
$ g(x) $
    via free quantum fields and zero modes like this
\begin{equation}
\label{lnf}
    \ln f(x) = Q - \frac{Px}{\pi} + \chi(x) ,
\end{equation}
    where 
$ Q $ and
$ P $ are canonical variables, whereas the periodic field
$ \chi(x) $
    is expressed by oscillators.
    The field
$ g(x) $ 
    is parametrized analogously with the same
$ P $ and
$ Q $.
    There arises a picture of the corresponding Hilbert space
\begin{equation*}
    \Hh = L_{2}(\Rr) \otimes \HH_{L} \otimes \HH_{R},
\end{equation*}
    where
$ \HH_{L} $ and
$ \HH_{R} $
    represents oscillators and
$ L_{2}(\Rr) $
    is a space for zero modes
$ (P,Q) $.
    However on some stage an invariance to reflection
$ P \to -P $ appears,
    the origin of which is usually quite vague.
    In this paper, which follows the previous publication
\cite{FV},
    I shall present arguments for this invariance and describe its realization
    in Hilbert space.

\section{Classical theory}
    The equation
(\ref{Leqm})
    is a condition of zero curvature
\begin{equation*}
    [\hat{L}_{1}, \hat{L}_{0}] = 0
\end{equation*}
    for the connection
\begin{equation*}
    \hat{L}_{1} = \frac{d}{dx} - L_{1} , \quad
    \hat{L}_{0} = \frac{d}{dt} - L_{0} , 
\end{equation*}
    where
$ 2\times 2 $
    matrices
$ L_{1} $ and
$ L_{0} $
    are parametrized by the field
$ \phi(x,t) $
\begin{equation*}
    L_{1} = \frac{1}{2} \begin{pmatrix}
	    \phi_{t} & e^{\phi} \\
	    e^{\phi} & -\phi_{t}
	\end{pmatrix} , \quad
    L_{0} = \frac{1}{2} \begin{pmatrix}
	    \phi_{x} & -e^{\phi} \\
	    e^{\phi} & -\phi_{x}
	\end{pmatrix} .
\end{equation*}

    Let
\begin{equation*}
    T(x,t) =
\begin{pmatrix}
    A(x,t) & B(x,t) \\
    C(x,t) & D(x,t)
\end{pmatrix}
\end{equation*}
    be the corresponding holonomy, i.e. a solution of the compatible
    equations
\begin{equation*}
    T_{x} = L_{1} T, \quad T_{t} = L_{0} T    
\end{equation*}
    with initial condition
\begin{equation}
\label{icon}
    T(0,0) = I .
\end{equation}
    It is easy to see, tha the projective components
\begin{equation*}
    f(x,t) = \frac{A(x,t)}{B(x,t)} , \quad g(x,t) = \frac{C(x,t)}{D(x,t)}
\end{equation*}
    satisfy the equations
\begin{equation*}
    f_{x} = f_{t} , \quad g_{x} = -g_{t}
\end{equation*}
    and thus
\begin{equation*}
    f(x,t) = f(x-t) , \quad g(x,t) = g(x+t) .
\end{equation*}
    It follows from
(\ref{icon}),
    that the following initial condition
\begin{equation*}
    f(0) = \infty , \quad g(0) = 0
\end{equation*}
    and bounds
\begin{equation*}
    f'(0) <0 , \quad g'(0) >0 
\end{equation*}
    are valid.
    It follows, that
$ f(x) $ and
$ g(x) $
    are positive.
    They realize the Liouville formula
(\ref{Lf}).

    Let
$ M $
    be a corresponding monodromy for a circle 
$ t=0 $
\begin{equation*}
    M = T(2\pi, 0) =
	\begin{pmatrix}
	    a & b \\
	    c & d
	\end{pmatrix} .
\end{equation*}
    It is a hyperbolic unimodular matrix with positive matrix elements.
    It is clear, that
$ f $ and
$ g $
    satisfy the condition of quasiperiodicity
\begin{equation*}
    f(x+2\pi) = \frac{af(x)+c}{bf(x)+d} , \quad
    g(x+2\pi) = \frac{ag(x)+c}{bg(x)+d} .
\end{equation*}
    These conditions simplify if one diagonalizes the monodromy taking
\begin{equation*}
    MN = ND ,
\end{equation*}
    where
\begin{equation*}
    D = \begin{pmatrix}
	    \lambda & 0 \\
	    0 & 1/\lambda
	\end{pmatrix} , \quad 
    N = \begin{pmatrix}
	    1 & 1/\xi \\
	    \eta & 1
	\end{pmatrix}
\end{equation*}
    The parameter
$ \lambda $
    defines the eigenvalues of monodromy and
$ \xi $,
$ \eta $
    are its fixed points, i.e. solutions of the quasiperiodic equation
\begin{equation}
\label{sqeq}
    b \xi^{2} - (a-d) \xi - c =0, \quad a+d = \lambda + \frac{1}{\lambda} .
\end{equation}
    Leon Takhtajan and me in 1984 calculated the Poisson brackets for
    matrix elements
$ a $, $ b $, $ c $, $ d $
    of the monodromy
$ M $
(published in 1986 in \cite{FT})
\begin{gather*}
    \{a,b\} = \frac{1}{2} \gamma ab , \quad \{a,c\} = \frac{1}{2}\gamma ac , \\
    \{d,b\} = -\frac{1}{2}\gamma bd , \quad \{d,c\} = -\frac{1}{2}\gamma cd ,\\
    \{b,c\} = 0 , \quad \{a,d\} = \gamma bc .
\end{gather*}
    It is instructive to note that these relations became a base
    of quantum groups
\cite{Drinfeld}, \cite{RTF}.

    It followes from these relations, that
$ \lambda$, $ \xi  $ and $ \eta $
    can be parametrized in the form
\begin{equation*}
    \lambda = e^{-P} , \quad \xi = \alpha e^{Q} , \quad \eta = -\alpha e^{-Q}
\end{equation*}
    with brackets
\begin{equation*}
    \{P,Q\} = \frac{\gamma}{2}, \quad \{P,\alpha\} = 0, \quad \{Q,\alpha\} =0 .
\end{equation*}
    In term of these variables the monodoromy 
$ T $
    assumes the form
\begin{equation}
\label{MT}
    M = \frac{1}{\ch Q} \begin{pmatrix}
	\ch(Q-P) & \alpha^{-1} \sh P \\
	\alpha \sh P & \ch(Q+P)
    \end{pmatrix} .
\end{equation}

    Now for the M\"obius transformed
$ f $
\begin{equation*}
    \hat{f} = N(f) = \frac{f+\eta}{f/\xi+1}
\end{equation*}
    we have
\begin{equation*}
    \hat{f}(0) = \xi = \alpha e^{Q} , \quad \hat{f}(2\pi) = e^{-2P} \hat{f}(0),
\end{equation*}
    so that
\begin{equation*}
    \ln \hat{f}(x) /\alpha = Q - \frac{Px}{\pi} + \chi(x) ,
\end{equation*}
    where
$ \chi(x) $ is periodic.
    We get the formula
(\ref{lnf}).
    However the elements of monodormy
$ M $
    are positive only under condition
\begin{equation*}
    P > 0 .
\end{equation*}
    Thus the phase space of the zero modes 
$ P $,
$ Q $
    is a halfplane, which raises the problem of correct quantization.
    We shall do it by initial choice of alternative canonical variables
    and appropriate canonical transformation.

    Introduce change of variables
\begin{equation*}
    u = \frac{\ch Q}{\ch(Q-P)} = \frac{1}{a} , \quad
    v = \frac{\sh^{2}P}{\ch Q \ch(Q-P)} = \frac{bc}{a} ,
\end{equation*}
    which maps the upper halfplane
$ (P,Q) $
    into the positive quadrant
\begin{equation*}
    u > 0 , \quad v > 0 .
\end{equation*}
    The brackets of
$ u $ and
$ v $
    are as follows
\begin{equation*}
    \{u,v\} = -\gamma uv ,
\end{equation*}
    so that their logarithms are canonical variables in
$ \Rr^{2} $
    and their quantization is trivial.

    The monodromy in new variables assumes the form
\begin{equation*}
    M = \begin{pmatrix}
	u^{-1} & \alpha \sqrt{v/u} \\
	\alpha^{-1} \sqrt{v/u} & u+v
    \end{pmatrix}
\end{equation*}
    and now we can construct its quantum realization.

\section{Quantization}
    Quantum analogue of the variables
$ u $ and
$ v $
    is a Weyl pair
$ u $ and
$ v $
    with relation
\begin{equation*}
    uv = q^{2} vu , \quad q = e^{i\gamma/2} .
\end{equation*}
    The variable
$ \alpha $
    remains a central element. To be closer to the notation of the theory
    of automorphic functions we put
\begin{equation*}
    \gamma = 2\pi \tau
\end{equation*}
    and parametrize
$ \tau $
    via imaginary halfperiods
$ \omega $, 
$ \omega' $
\begin{equation*}
    \tau = \frac{\omega'}{\omega} , \quad \omega \omega' = -\frac{1}{4} ,
\end{equation*}
    lying in the upper halfplane.

    The operators
$ u $ and $ v $
    can be realized in
$ L_{2}(\Rr) $
    in the form
\begin{equation*}
    u f(x) = e^{-i\pi x/\omega} f(x) , \quad v f(x) = f(x+2\omega') 
\end{equation*}
    as essentially selfadjoint operators with the domain
$ \Dd $,
    consisting of analytic functions like
\begin{equation*}
    e^{-\alpha x^{2}} e^{\beta x} P(x) ,
\end{equation*}
    with
$ \alpha >0 $,
$ \beta \in \CC $ and
    $ P(x) $ being a polynomial with complex coefficients.

    Quantum analogue of the monodromy is given by matrix
\begin{equation*}
    M = \begin{pmatrix}
	u^{-1} & \alpha q^{-1/4} u^{-1/2} v^{1/2} \\
	\alpha^{-1} q^{-1/4} u^{-1/2} v^{1/2} & u+v
    \end{pmatrix} 
\end{equation*}
    after the natural ordering of the factors.

    Our first problem is to diagonalize
$ \tr M $, namely the operator
\begin{equation*}
    L = u+u^{-1} +v .
\end{equation*}
    This operator is very well known nowadays and it acquired a lot of
    interpretations. For exmple in the quantum Teichmuller theory
\cite{Kashaev1}, \cite{ChF}
    it is called operator of minimal geodesic length,
    in the theory of the decomposition od the tensor product of two
    irreducible representations of modular double
$ SL_{q}(2,\Rr) $
    into irreducibles it appears as a main part of corresponding Casimir
\cite{DF}.
    Spectral theory for
$ L $
    was investigated by Kashaev
\cite{Kashaev2}.

    Operator
$ L $
    has continuos spectrum of multiplicity 1 on the halfline
$ 2 <\lambda < \infty $.
    The corresponding generalized eigenfunctions can be expressed
    in terms of function
\begin{equation*}
    \gamma(x) = \exp -\frac{1}{4} \int_{-\infty}^{\infty}
	\frac{e^{ixt}}{\sin\omega t \, \sin \omega' t} \frac{dt}{t} ,
\end{equation*}
    where the singularity at
$ t=0 $
    is passed from above.
    This function acquired recently quite a popularity.
    I call it ``quantum modular dilogarithm''
\cite{Faddeev}.
    Detailed description of its properties is given in
\cite{Volkov}.
    Those of them, used in this paper, are collected in Appendix.

    The basic property --- the functional equation
\begin{equation}
\label{feq}
    \frac{\gamma(x+\omega')}{\gamma(x-\omega')} = 1+ e^{-i\pi x/\omega}
\end{equation}
    can be used to show, that the generalized function
\begin{equation*}
    \psi(x,s) = \gamma(x-s-\omega'' +i0) \gamma(x+s-\omega''+i0)
	e^{-i\pi(x-\omega'')^{2}}
\end{equation*}
    for real
$ s $
    is an eigenfunction of operator
$ L $
    with eigenvalue
\begin{equation*}
    e^{i\pi s/\omega} + e^{-i\pi s/\omega} = 2 \cos \pi s/\omega ,
\end{equation*}
    where
\begin{equation*}
    \omega'' = \omega + \omega' .
\end{equation*}
    Function
$ \gamma(z-\omega'') $
    has a pole at
$ z=0 $
    and
$ +i0 $
    is added in the expression for
$ \psi(x,s) $
    to define how it is understood.

    Kashaev
\cite{Kashaev2}
    proved orthonormality of functions
$ \psi(x,s) $
    in the form
\begin{align*}
    \int \overline{\psi(x,s)} \psi(x,s') dx = & \frac{1}{\rho(s)}
	\bigl( \delta(s-s') + \delta(s+s')\bigr) \\
    \int_{0}^{\infty} \psi(x,s) \overline{\psi(y,s)} \rho(s) ds & =
	\delta(x-y) ,
\end{align*}
    where the measure
$ \rho(s) $
    is given by
\begin{equation}
\label{mef}
    \rho(s) = -4\sin \frac{\pi s}{\omega} \, \sin \frac{\pi s}{\omega'} .
\end{equation}
    Thus the natural statement is as follows:
    the diagonal representation for operator
$ L $
    is Hilbert space
$ L_{2}(\Rr_{+},\rho) $.

    Let us calculate the expression for the nondiagonal element of monodromy
    in this representation.
    Denote by
$ U $
    an integral operator with kernel
$ \psi(x,s) $,
    which connects spaces
$ L_{2}(\Rr_{+},\rho) $ and
$ L_{2}(\Rr) $.
    Let
$ r $ and
$ Z $
    be a Weyl pair acting on functions
$ F(s) $
    for
$ s \in \Rr $
\begin{equation*}
    r F(s) = F(s+\omega') , \quad Zf = e^{i\pi s/\omega} F(s) .
\end{equation*}
    Later we shall find contributions of
$ r $ and
$ Z $
    having sense in
$ L_{2}(\Rr_{+},\rho) $.

    From the functional equation
(\ref{feq})
    we get the expressions for the kernels of integral operators
$ v^{1/2} Ur $ and
$ v^{1/2} Ur^{-1} $
\begin{align*}
    \psi(x+\omega', s-\omega') = & (1-\frac{u}{Z})\frac{1}{i} q^{-1/2} u^{-1/2}
	\psi(x,s) \\
    \psi(x+\omega', s+\omega') = & (1-uZ)\frac{1}{i} q^{-1/2} u^{-1/2}
	\psi(x,s) ,
\end{align*}
    and from this we get
\begin{equation*}
    q^{1/4} u^{-1/2} v^{1/2} U = U\frac{1}{i} (Z-Z^{-1})(r-r^{-1})^{-1} .
\end{equation*}
    Operator in the RHS makes sense in the subspace of even functions of
$ s $
    and is selfadjoint in
$ L_{2}(\Rr_{+},\rho) $.

    Thus we get a new representation for the matrix elements of monodromy
\begin{equation*}
    U^{-1} (a + d) U = Z+Z^{-1} , \quad
	U^{-1} bU = \frac{\alpha}{i} (Z-Z^{-1}) \frac{1}{r-r^{-1}} .
\end{equation*}
    It is instructive to compare these quantum frmulas with the classical
    ones in
(\ref{MT})
\begin{equation*}
    a+d = 2 \ch P , \quad b=\alpha \frac{\sh P}{\ch Q} .
\end{equation*}
    It is natural to identify
$ Z $
    with
$ e^{P} $, and
$ r^{-1} $ with
$ e^{Q} $,
    however classical
$ \ch Q $
    is changed into
$ i\sh Q $;
    this corresponds to shift
$ Q $ by
$ i\pi/2 $.

    I was not able to solve the quantum analogue of equation
(\ref{sqeq})
    to get the representations of
$ u $ and
$ v $.

\section{Reflection coefficient}
    It is natural to want to get rid of the measure
$ \rho(s) $
    and find the realization of operators
$ P $ and
$ b $
    in usual
$ L_{2}(\Rr) $.
    It is clear, that such a realization can be done only in a subspace.
    We shall show, that the corresponding projector looks like
\begin{equation*}
    \Pi = \frac{1}{2} (I+KS) ,
\end{equation*}
    where
$ K $ is a simple reflection
\begin{equation*}
    K F(s) = F(-s) ,
\end{equation*}
    and
$ S $
    is an operator of multiplication by a function
$ S(s) $
    such that
\begin{equation*}
    \overline{S(s)} = S(-s) = S^{-1}(s) .
\end{equation*}
    We call this factor the reflection coefficient.

    The explicite calculation of
$ S(s) $
    is based on factorization
\begin{equation*}
    1/\rho(s) = M(s) M(-s) ,
\end{equation*}
    where
\begin{equation*}
    M(s) = \text{const} e^{-2i\pi s(s-\omega'')} \gamma(2s -\omega''),
\end{equation*}
    with constant which garantees the relation
\begin{equation*}
    \overline{M(s)} = M(-s) ,
\end{equation*}
    with the help of relations
(\ref{A1}), (\ref{A3})
    from the Appendix. The explicite value of this constant is inessential
    for us in what follows.

    From the functional relation
(\ref{feq})
    it follows
\begin{align}
\label{Mplus}
    M(s+\omega') = & M(s) \frac{1}{i} (Z-Z^{-1}) \\
\label{Mminus}
    M(s-\omega') = & M(s) i (q^{-1}Z-qZ^{-1})^{-1} .
\end{align}
    Let us define an operator
\begin{equation*}
    V = U \frac{1}{M(s)} .
\end{equation*}
    The image of
$ V $
    consists of functions
$ F(s) $
    satisfying condition
\begin{equation*}
    F(-s) = S(s) F(s) ,
\end{equation*}
    where
\begin{equation*}
    S(s) = \frac{M(s)}{M(-s)},
\end{equation*}
    and thus belonging to subspace
\begin{equation*}
    \Pi L_{2}(\Rr) ,
\end{equation*}
    which we shall take as a natural Hilbert space for the zero mode
$ P $.

    Operator
$ b^{-1} = q^{1/4} u^{1/2} v^{-1/2} $
    is transfrmed by
$ V $
    as follows
\begin{align*}
    b^{-1} V = & U(r-r^{-1}) (Z-Z^{-1})^{-1} \frac{1}{M(s)} = \\
	= & V\bigl( r^{-1} + \frac{1}{Z-Z^{-1}} r \frac{1}{Z-Z^{-1}}\bigr) .
\end{align*}
    The formula in the RHS was proposed in
\cite{FV}
    using the quasiclassical considerations. Here we get its quantum
    derivation.

    To conclude: we have defined a natural quantum Hilbert space for the
    zero modes
$ P $ and
$ Q $.

    In papers
\cite{ZZ}, \cite{JW}
    the alternative realization was proposed with the reflection
    coefficient expressed in terms of classical
$ \Gamma $-function.
    The factorization of the measure
$ \rho(s) $
(\ref{mef})
    makes sense also fr this reflection coefficient.
    It will be interesting to make a more close contact between the two
    approaches.

\section{Evolution}
    In 
\cite{FV}
    it was shown, that under discrete shift of time
\begin{equation*}
    t \to t +\pi
\end{equation*}
    the zero modes
$ P $ and
$ Q $
    transforms elementary
\begin{equation*}
    P \to P , \quad Q \to P+Q .
\end{equation*}
    This is connected with the fact that this evolution is product of those
    for left and right motion
%\begin{equation*}
%    t \to t+\pi , \xi \to \xi -\pi , \quad
%    t \to t+\pi , \xi \to \xi +\pi 
%\end{equation*}
    and oscillator degrees of freedom do not change.

    For the variables
$ u $ and
$ v $
    this evolution is given by
\begin{align*}
    u \to & u+v \\
    v \to & u^{-1} v (u+v)^{-1} .
\end{align*}
    We shall assume that this is true also in quantum case with already
    prescribed order of factors.
    We look for the corresponding evolution operator
\begin{equation*}
    K^{-1} u K = u+v , \quad K^{-1} v K = u^{-1} v(u+v)^{-1} .
\end{equation*}
    It is easier to find the operator
$ K^{-1} $
\begin{equation*}
    (u+v) K^{-1} = K^{-1} u , \quad K u^{-1}v K^{-1} = K^{-1} vu ,
\end{equation*}
    where we simplified the second equation with proper order of factors.

    It follows from the function equation
(\ref{feq})
    that
$ K^{-1} $
    is an integrable operator with kernel
\begin{equation*}
    K^{-1}(x,y) = e^{2\pi ixy} \gamma(x-y-\omega'') .
\end{equation*}
    Let us show, that
$ \psi(x,s) $
    is its eigenfunction.
    Consider
\begin{align*}
    K^{-1}\psi =& \int e^{2\pi ixy} \gamma(x-y-\omega'') \gamma(y+s-\omega'')
	\gamma(y-s-\omega'') e^{-i\pi(y-\omega'')^{2}} dy =\\
    =& I(x,s) .
\end{align*}
    The integral in the RHS can be explicitely calculated with the use of a
    particular variant of so called
$ \beta $-integral (see e.g.
\cite{Volkov}).

    Begin by rearranging the expression for
$ \psi(x,s) $
    using the reflection property
\begin{equation*}
    \psi(x,s) = c(s) \frac{\gamma(x+s-\omega'')}{\gamma(s-x+\omega'')}
	e^{2\pi isx},
\end{equation*}
    where
\begin{equation*}
    c(s) = \beta e^{-i\pi s^{2}} .
\end{equation*}
    Then use the integral identity
(\ref{A5})
\begin{equation*}
    \frac{\gamma(x-y-\omega'')}{\gamma(s-y+\omega'')} =
	\frac{c}{\gamma(s-x+\omega'')}
    \int\frac{\gamma(s-x+\omega''+z)}{\gamma(z+\omega'')} e^{2\pi iz(x-y)} dz.
\end{equation*}
    for two factors in the integral
$ I(x,s) $ and formula
(\ref{A7})
    for the integral of the last factor
\begin{equation*}
    \int \gamma(y+s-\omega'') e^{2\pi i(x-z-s)} dy
	= c \frac{1}{\gamma(z-x+s+\omega'')} .
\end{equation*}
    Factors
$ \gamma(s-x+\omega''-z) $
    cancel and as result
\begin{equation*}
    I(x,s) = c^{2} c(s) \frac{e^{2\pi is^{2}}}{\gamma(s-x+\omega'')} 
	\int \frac{e^{2\pi iz(x+s)}}{\gamma(z+\omega'')} dz
	= c^{2} e^{2\pi is^{2}} \psi(x,s) ,
\end{equation*}
    with the help of
(\ref{A6}).

    Thus Kashaev function
$ \psi(x,s) $
    is an eigenfunction of operator
$ K^{-1} $
    with the eigenvalue
$ c^{2} e^{2\pi is^{2}} $.
    The constant factor
$ c^{2} $
    can be omitted, so that the evolution operator
$ K $
    in ur representation is simply the multiplication by
$ e^{-2\pi is^{2}} $
    so that
\begin{equation*}
    K^{-1}ZK = Z , \quad K^{-1}rK
	= e^{2\pi is^{2}} e^{-2\pi i(s+\omega')^{2}} r
	= q^{1/2} e^{i\pi s/\omega} r = q^{1/2} Zr.
\end{equation*}
    Exactly this evolution was obtained in
\cite{FV}.

\section{Conclusion}
    It is shown, that the natural inclusion of the zero modes of the
    Liouville model into the monodromy matrix of the Lax operator
$ \hat{L}_{1} $
    gives the description of the Hilbert space and the evolution operator
    for the quasimomentum
$ P $.

\section{Appendix}
    The properties of the modular quantum dilogarithm are given in detail
    in the paper
\cite{Volkov}
    as well as in many other places. In this paper we used the following
    properties besides the fundamental equation
(\ref{feq}):

    The reflection condition
\begin{equation}
\label{A1}
    \gamma(z) \gamma(-z) = \beta e^{i\pi z^{2}} , \quad
	\beta = e^{\frac{i\pi}{12}(\tau+\frac{1}{\tau})} .
\end{equation}

    The description of the first pole
\begin{equation}
\label{A2}
    \gamma(z+\omega'') = \frac{c}{z} , \quad
	c = -\frac{1}{2\pi i\beta} e^{-i\pi/4} 
\end{equation}

    ``Reality''
\begin{equation}
\label{A3}
    \overline{\gamma(z)} = \frac{1}{\gamma(\bar{z})}
\end{equation}

    Asymptotics
\begin{equation}
\label{A4}
    \gamma(z) \to 1
\end{equation}
    in the sector
$ -\frac{\pi}{4} < \arg z < \frac{\pi}{4} $.

    Integral identity
\begin{equation}
\label{A5}
    \frac{\gamma(t+a)}{\gamma(t+b)} = \frac{c}{\gamma(b-a-\omega'')}
	\int \frac{\gamma(b-a-\omega''+z)}{\gamma(z+\omega'')}
	    e^{2\pi iz(t+a+\omega'')} dz ,
\end{equation}
    its limit for
$ b \to \infty $
\begin{equation}
\label{A6}
    \gamma(t+a) =
	c \int \frac{e^{2\pi iz(t+a+\omega'')}}{\gamma(z+\omega'')} dz
\end{equation}
    and its inversion
\begin{equation}
\label{A7}
    \int \gamma(t+a) e^{-2\pi its} dt =
	c \frac{e^{2\pi is(a+\omega'')}}{\gamma(s+\omega'')} .
\end{equation}

\end{document}